\documentclass[prl,twocolumn,superscriptaddress,showpacs,a4paper]{revtex4}

\usepackage[dvips]{graphicx}

\begin{document}


\title{Self-Consistent Scaling Theory for Logarithmic Correction Exponents}
\date{September 2006}

\author{R.\ Kenna}
\affiliation{Applied Mathematics Research Centre,
Coventry University, Coventry, CV1 5FB, England}

\author{D.A.\ Johnston}
\affiliation{Department of Mathematics and the Maxwell Institute for Mathematical Sciences,
Heriot-Watt University, Riccarton, Edinburgh, EH14 4AS, Scotland}

\author{W.\ Janke}
\affiliation{Institut f\"ur Theoretische Physik and Centre for Theoretical Sciences (NTZ), Universit\"at Leipzig,
Augustusplatz 10/11, 04109 Leipzig, Germany}

\begin{abstract}
Multiplicative logarithmic corrections frequently characterize critical behaviour in statistical physics.
Here, a recently proposed 
theory relating the exponents of such 
terms
is extended to account for circumstances which often occur when
the leading specific-heat critical exponent vanishes. 
Also, the  theory is widened to encompass the correlation function.
The new  relations are then confronted with results from the literature
and some new predictions for logarithmic corrections 
in certain models are made.
\end{abstract}



\maketitle


In a recent Letter, we presented three new relations between some of the exponents of multiplicative logarithmic corrections to scaling
which are frequently manifest in statistical physics \cite{I}.
While two of these relations were demonstrated to hold generally, the third fails in certain circumstances 
where the specific-heat leading exponent $\alpha$ vanishes.
Here, a broad theory which encompasses such scenarios is presented and a fourth general scaling relation for such logarithms is derived.
Together, these amount to logarithmic analogs of the standard scaling relations between the leading exponents,
which are well established and fundamentally important in statistical mechanics (see e.g. Refs.~\cite{history,ItDr89} and references therein). 

Denoting the reduced temperature by $t$, we address the circumstances in $d$ dimensions where the correlation length, specific heat, 
susceptibility and correlation function scale respectively as \cite{I}
\begin{eqnarray}
\xi_\infty (t) & \sim & |t|^{-\nu} |\ln{|t|}|^{\hat{\nu}}
\,,
\label{xi}
\\
  C_\infty (t) & \sim & |t|^{-\alpha} |\ln{|t|}|^{\hat{\alpha}}
\,,
\label{C}
\\
\chi_\infty (t) & \sim & |t|^{-\gamma} |\ln{|t|}|^{\hat{\gamma}}
\,,
\label{chi}
\\
{\cal{G}}_\infty (x,t) & \sim & x^{-(d-2+\eta)}(\ln{x})^{\hat{\eta}} 
 D\left(
 \frac{x}{\xi_\infty(t)}
 \right)
\,,
\label{corrfun}
\end{eqnarray}
in which  $x$ represents position on the lattice, whose extent is  indicated by the subscript.
When this is finite, the counterpart of (\ref{xi}) 
may be assumed to be  \cite{I}
\begin{equation}
 \xi_L(0) \sim L (\ln{L})^{\hat{q}}
\,.
\label{corrL}
\end{equation}

The aim of this Letter is to establish full logarithmic analogs of the following two standard scaling relations;
\begin{eqnarray}
 \nu d & = & 2-\alpha 
\,,
\label{Jo}
\\
 \nu (2-\eta) &  = & \gamma 
\,.
\label{Fi}
\end{eqnarray}
The relation (\ref{Jo}) was developed by Widom \cite{Wi65} (see also Ref.~\cite{Gr67}) who also showed how  a logarithmic singularity
may arise in the specific heat if $\alpha=0$ with, in general, a superimposed finite discontinuity (see also Ref.~\cite{Fi65}).
The second relation (\ref{Fi}) is due to Fisher  \cite{FiSR64}.
For an authoritative and comprehensive outline of the development of the original scaling relations  the reader is referred to Ref.~\cite{history}.

The scaling theory presented herein is based on self-consistencies, which are manifest as relations between the various
correction exponents.
For ab inito model-specific theories,  the renormalization group and related approaches are appropriate
\cite{approp} and the reader is again referred to Ref.~\cite{history} for a review.  
Our approach is not dependent on such renormalization group considerations.

In Ref.~\cite{I}, we used a Lee-Yang analysis
to establish the following scaling relation between the exponents of the logarithmic corrections analogous to (\ref{Jo});
\begin{equation}
  \hat{\alpha} =  d \hat{q} -  d \hat{\nu}
\,,
\label{NewK}
\end{equation}
and this formula was confronted with a 
variety of results from the literature.
While it holds in most models, exceptional cases that were identified include  the pure Ising model in two
dimensions and its  uncorrelated, quenched,  random disordered counterpart.
Indeed, it is not altogether surprising that a Lee-Yang analysis, which focuses on a complex odd (magnetic) scaling field,
cannot completely realize the general relationship between the even correction exponents appearing in (\ref{NewK}).
The first main aim of this Letter is to redress this situation by appealing 
to Fisher zeros, which are appropriate to the 
even sector and to present a complete theory for the logarithmic analog of (\ref{Jo}), which
also neatly encapsulates the $d=2$ (pure and random) Ising cases. 
The second main aim 
is to present a complementary analog of Fisher's scaling relation (\ref{Fi}) involving the correlation-function correction 
exponent $\hat{\eta}$ and confront it with the literature.
We now address these two issues in sequence.

In two dimensions, the pure Ising model has $\hat{\alpha} = 1$ and, since all other correction exponents vanish,
 (\ref{NewK}) fails there.
It also fails in the version with random-bond  disorder, where ${\hat{q}} = 0$  \cite{Aade96,LaIg00}, 
$\hat{\alpha} =0 $, $\hat{\nu}   = 1/2  $  and
\cite{SSLJ,DD}
\begin{equation}
C_\infty(t) \sim \ln{|\ln{|t|}|}
\,.
\label{dL}
\end{equation}
Numerical works supportive of the vanishing of $\hat{\alpha}$ and the double-logarithmically divergent
specific heat are found in Refs.~\cite{Aade96,AnDo90WaSe90,TaSh94WiDo95StAa97RoAd98} (see also Ref.~\cite{BeSh04BeCh04})
and  Ref.~\cite{BaFe97SeSh98Pl98} for both the
random-bond and site-diluted 
models, respectively.
However, 
there has been considerable disparity in the literature as to the
precise scaling behavior
and
counter claims that the specific heat remains finite
in the random-bond \cite{Ki00,finiteCinBond} and random-site 
versions \cite{ZieglerHe92,KiPa94Ku94KuMa00} also exist
(see also Ref.~\cite{Se94Zi94vaKu00}).

While it was  mooted in Ref.~\cite{I} that the detailed logarithmic corrections 
in the random-bond and random-site Ising models in $d=2$ dimensions 
may in principle differ, it is herein clarified that this is not, in fact, 
expected to be the case.
Whereas in Ref.~\cite{I}, the Lee-Yang zeros were used to link the even and odd scaling fields,
we now appeal to the Fisher zeros of the even sector \cite{Fi65}, 
as that is where the apparent specific heat
anomaly related to (\ref{NewK}) lies.
The puzzle is resolved as being due to 
two special properties of the pure and random Ising models,
namely the vanishing of $\alpha$ and the manner in which the Fisher zeros in 
these models impact onto the real axis.

From the finite-size scaling (FSS) hypothesis, one has, 
for the  specific heat \cite{KeLa91},
\begin{equation}
 \frac{C_L(0)}{C_{\infty}(t)}
 =
 {\cal{F_C}} 
 \left(
         \frac{\xi_L(0)}{\xi_\infty(t)}
 \right)
\,.
\label{modFSSC}
\end{equation}
Fixing the scaling ratio $\xi_L(0)/\xi_\infty(t)$ gives
$t \sim L^{-1/\nu}(\ln{L})^{(\hat{\nu}-\hat{q})/\nu}$, 
which from (\ref{C}) yields
\begin{equation}
 C_L(0)
 \sim
 L^{\frac{\alpha}{\nu}}
 (\ln{L})^{\hat{\alpha} - \alpha \frac{\hat{\nu}-\hat{q}}{\nu}}
\,.
\label{B}
\end{equation}
A FSS theory for partition function zeros for pure power-law scaling 
was formulated in Ref.~\cite{IPZ}
by writing the partition function for a finite-size system as a function of the scaling ratio there.
Here, allowing for logarithmic corrections, this partition function may be written as
$Z_L(t) = Q\left( {\xi_L(0)}/{\xi_\infty(t)} \right)$
and  vanishes at a Fisher zero. Labeling the $j^{\rm{th}}$ such zero
as $t_j(L)$, one has
\begin{equation}
 \frac{\xi_L(0)}{\xi_\infty(t_j(L))} = Q_j^{-1}(0)
\,.
\end{equation}
where $Q_j^{-1}(0)$ is the $j^{\rm{th}}$ complex root of $Q$. Therefore
\begin{equation}
 |t_j(L)| \sim L^{-\frac{1}{\nu}}
 (\ln{L})^{
            \frac{ \hat{\nu}-\hat{q} }{\nu}}
\,.
\label{A}
\end{equation}
No assumptions other than the validity of
FSS  have been used to derive  (\ref{B}) and (\ref{A}).

The total number of conjugate pairs of zeros, $\cal{N}$, in a suitable variable $t$ is proportional
to the lattice volume so that ${\cal{N}}\propto L^d$. 
The full expression for the scaling of the $j^{\rm{th}}$ zero is given in 
Ref.~\cite{JaKe01} (see also Refs.~\cite{IPZ,GPS}) as a function of a
fraction of the total number of
zeros $(2j-1)/2L^d$. 
Then, allowing for logarithmic corrections, 
(\ref{A}) is more appropriately 
written as
\begin{equation}
 t_j(L) \sim
 \left(\frac{j-1/2}{L^d}
 \right)^{\frac{1}{\nu d}}
 \left(
    \ln{\left( \frac{j-1/2}{L^d} \right)}
 \right)^\frac{\hat{\nu}-\hat{q}}{\nu}
 \exp{(i \phi_j(L))}
\,,
\label{most}
\end{equation}
where $\phi_j(L)$ is the argument of the $j$th zero.
In all known cases,  the Fisher zeros for isotropic models on homopolygonal lattices  
lie on curves
in the complex plane and impact onto the real axis 
along a singular line \cite{MaSh95}.
We assume this scenario, and denote the impact angle 
onto the real axis in the thermodynamic limit by  $\phi$.

Now, writing the finite-size partition function in terms of its Fisher zeros or free-energy singularities,
\begin{equation}
 Z_L(t) \propto \prod_{j=1}^{\cal{N}} \left(t - t_j(L)\right)\left(t - t^*_j(L)\right)
\,,
\end{equation}
where $t_j(L)$ and $t^*_j(L)$ are complex conjugate pairs.
Assume that the
${\cal{M}} \propto {\cal{N}}$ 
zeros 
which dominate  scaling behavior close to the critical point 
are described by the scaling form (\ref{most}).
Appropriate differentiation gives for the specific heat at $t=0$
\begin{equation}
 C_L(0) \sim - L^{-d} {\rm{Re}}\sum_{j=1}^{\cal{M}} t_j^{-2}(L)
\,,
\label{Cc}
\end{equation}
having included the volume factor $L^{-d}$. 

 In the case where $\nu d \ne 2$, so that $\alpha \ne 0$ by (\ref{Jo}), one finds 
that 
the FSS expression (\ref{Cc}) gives for the singular part of the specific heat
\begin{equation}
 C_L(0) \sim L^{-d+\frac{\nu}{2}}  \left(
    \ln{L} 
 \right)^{-2\frac{\hat{\nu}-\hat{q}}{\nu}}
\,.
\label{D}
\end{equation}
Comparison with ({\ref{B}) leads to the recovery of~(\ref{NewK}).

If, however, $\alpha = 0$, so that $\nu d = 2$ by (\ref{Jo}), the FSS expression (\ref{Cc}) 
for the specific heat becomes
\begin{equation}
 C_L(0) \sim
 \sum_{j=1}^{\cal{M}}{ 
 \frac{\cos{(2\phi_j(L))}}{j-1/2}
 \left(
    \ln{\left( \frac{j-1/2}{L^d} \right)}
 \right)^{-2 \frac{\hat{\nu}-\hat{q}}{\nu}}
}
\,.
\label{enrage}
\end{equation}
For sufficiently large $L$ and close to the transition point, 
$\phi_j(L) \simeq \phi$ and the cosine term in (\ref{enrage}) becomes a non-zero constant
provided $\phi \ne \pi/4$.
This is the case in the  square-lattice pure Ising model in $d=2$ dimensions, 
where $\phi = \pi/2$  \cite{Fi65}.
Simple invariance symmetries (such as self-duality or 
duality combined with the star-triangle relation) 
which the distribution of Fisher zeros must respect
ensure that this is also the case with the pure model on other lattices \cite{symm}
as well as for the symmetric random-bond counterpart \cite{Fi78}. 
On continuity grounds, one also expects $\phi \ne \pi/4$  
in the general random-bond and random-site Ising models in two dimensions.

In these cases, from the Euler-Maclaurin formula,
the leading scaling behavior for large $L$ when $\alpha=0$ is 
\begin{equation}
 C_L(0)
 \sim 
 \left\{ 
        \begin{array}{ll}
                         (\ln{L})^{
                                   1-2 \frac{ \hat{\nu}-\hat{q} }{ \nu }
                                  } & \mbox{
                                            if $2(\hat{\nu}-\hat{q}) \ne \nu$
                                           } \\
                        \ln{\ln{L}} & \mbox{
                                            if $2(\hat{\nu}-\hat{q}) = \nu$
                                            } \,.
        \end{array}
 \right.
\label{doublelog}
\end{equation}
In the thermodyamic limit, $|t|$ and $C_\infty(t)$  replace $L$ and $C_L(0)$ in (\ref{doublelog}), respectively.
Comparing  (\ref{B}) with (\ref{doublelog}) and using (\ref{Jo}), one finds
\begin{equation}
 \hat{\alpha}
 = 
 1 + d \hat{q} - d\hat{\nu} 
\,.
\label{NewK2}
\end{equation}
This formula replaces (\ref{NewK}) in such circumstances where the model has 
 $\alpha = 0$ and $\phi \ne \pi/4$.
In the  pure Ising model in $d=2$ dimensions, where $\hat{q}=\hat{\nu}=0$,
(\ref{NewK2}) gives, correctly,  $\hat{\alpha}=1$ and the divergence of the 
specific heat there is caused by the extra logarithm as compared with (\ref{NewK})
(see also Refs.~\cite{Wi65,Gr67,Fi65}).
In the random $d=2$ Ising model where $\hat{q}=0$, $\hat{\nu}=1/2$, it gives $\hat{\alpha}=0$.
In general, if $2(\hat{\nu}-\hat{q}) = \nu$, 
the specific heat instead diverges with a double logarithm after (\ref{doublelog}).
This is precisely the circumstances in the random Ising model
in two dimensions \cite{SSLJ,DD}.

The $N$-colour Ashkin-Teller model also has $\alpha=0$ and is self-dual \cite{GrWi81}
with 
$\hat{\alpha} = -N/(N-2)$ and
$\hat{\nu}   = (N-1)/(N-2)  $
\cite{SSLJ}.
If $\hat{q}=0$, these values also support the new scaling relation (\ref{NewK2}).

The $O(N)$ symmetric $\phi^4$ theories (with short- or long-range interactions) 
at their upper critical dimension also have $\alpha=0$. 
There, however, $\phi=\pi/4$ \cite{IPZ,GPS,KeLa91}
so that (\ref{NewK2}) does not follow from (\ref{enrage}).
Instead
(\ref{NewK}) remains valid there as demonstrated in Ref.~\cite{I}.

We now turn our attention to the correlation function (\ref{corrfun}) and
a new scaling relation for $\hat{\eta}$, analogous to  (\ref{Fi}).
Firstly, fixing the argument of the function $D$ in (\ref{corrfun}), 
\begin{equation}
 {\cal{G}}_\infty (x,t)  \sim 
 \xi_\infty(t)^{-(d-2+\eta)}(\ln{\xi_\infty(t)})^{\hat{\eta}} 
 D\left( \frac{x}{\xi_\infty(t)} \right)
\,.
\label{corrfun1}
\end{equation}
Following Ref.~\cite{ItDr89} for example, and writing the 
singular part of the magnetic susceptibility as
\begin{equation}
 \chi_\infty(t) = \int_0^{\xi_\infty(t)}{d^dx {\cal{G}}_\infty (x,t)}
\,,
\end{equation}
one obtains
\begin{equation}
 \chi_\infty(t) \sim \xi_\infty(t)^{2-\eta} (\ln{\xi_\infty(t)})^{\hat{\eta}}
\,.
\end{equation}
From (\ref{xi}) and (\ref{chi}), the leading scaling recovers (\ref{Fi}).
Matching the logarithmic corrections yields
\begin{equation}
 \hat{\eta} =  \hat{\gamma} - \hat{\nu} (2 - \eta)
\,.
\label{red}
\end{equation}
This approach, obtaining the susceptibility from the correlation function, 
comes from the original one used by Fisher \cite{FiSR64} and has
also been used in Ref.~\cite{SaSo97}
for the $d=2$ four-state Potts model.
In fact, there $\eta = 1/4$, 
$\hat{\gamma} = 3/4$, $\hat{\nu} = 1/2$, $\hat{\eta} = -1/8$ \cite{leadPNaSc80CaNa80,SaSo97}
and 
(\ref{red}) holds.

For average quantities in the random Ising models in $d=2$ dimensions, 
$\eta = 1/4$, $\hat{\gamma} = 7/8$,
$\hat{\nu}   = 1/2$ and $\hat{\eta}=0 $ \cite{SSLJ}
and 
(\ref{red}) is again obeyed.
This value for $\hat{\eta}$ has been convincingly verified  numerically
\cite{KiPa94Ku94KuMa00,Ki00,MaKu99,AnDo90WaSe90,moreetahat}.
The new relation (\ref{red}) also holds in the $N$-colour Ashkin-Teller model,
which, along with  
$\eta = 1/4$,
$\hat{\gamma} = 7(N-1)/4(N-2)$,
$\hat{\nu}   = (N-1)/(N-2)  $,
has
$\hat{\eta}   = 0  $  \cite{SSLJ}.

The $O(N)$ symmetric $\phi^4$ theories at their upper critical dimension  
$d=d_c=4$ have 
$\eta = 0$, $\hat{\gamma} = (N+2)/(N+8)$,
$\hat{\nu}   = (N+2)/2(N+8) $ and $\hat{\eta}   = 0 $  \cite{BLZ},
and the expression   (\ref{red}) is obeyed.
Likewise,  $O(N)$ spin models with long-range interactions
decaying as $x^{-(d+\sigma)}$ 
have logarithmic corrections at $d=d_c=2\sigma$. There,
$\eta = 2-\sigma$ \cite{FiMa72}, 
$\hat{\gamma} = (N+2)/(N+8)$,
$\hat{\nu}   = (N+2)/\sigma(N+8) $ and the relation
(\ref{red}) correctly yields $\hat{\eta}   = 0 $
\cite{LuBl97}.

For the percolation problem, 
$\eta = 0$,
$\hat{\gamma} = 2/7$, 
and 
$\hat{\nu}   = 5/42 $
at the upper critical dimension $d_c=6$ \cite{Pottslogs}.
The correction exponent for the correlation function there
has recently been calculated to be $\hat{\eta}   = 1/21 $ \cite{StJa03}.
Again, this set of values satisfies  (\ref{red}).

Finally, (\ref{red}) can be used to predict the value of $\hat{\eta}  $
in other models, such as $m$-component spin glasses and Yang-Lee edge 
problems
 at their upper critical dimension $d_c=6$.
 For the former,
$\eta = 0$ \cite{spinglass},
$\hat{\gamma} = 2m/(2m-1)$,
$\hat{\nu}   = 5m/6(2m-1) $ \cite{Ru98}
giving  $\hat{\eta}   = m/3(2m-1)$. 
For the Yang-Lee problem,
$\eta = 0$,
$\hat{\gamma} = 2/3$, 
$\hat{\nu}   = 5/18 $ \cite{Ru98}
so that the prediction from (\ref{red}) is 
  $\hat{\eta}   = 1/9$.
These values remain to be verified  numerically.

It is observed in Ref.~\cite{SaSo97}, that the magnetization $m_\infty(t)$
for the four-state Potts model may be deduced from the correlation function
by an alternative argument;
representing a generic spin-type variable by ${\vec{s}}(x)$,
if the spins decorrelate  in the limit where $x \rightarrow \infty$
 such that ${\cal{G}}_\infty (x,t) = \langle {\vec{s}}(0){\vec{s}}(x)\rangle
\rightarrow \langle {\vec{s}}(0)\rangle \langle {\vec{s}}(x)\rangle
= m_\infty^2(t)$ there, then
using (\ref{xi}) and (\ref{corrfun1}) and matching with $m_\infty(t) \sim |t|^\beta |\ln{|t|}|^{\hat{\beta}}$ \cite{I}
gives $\nu (d-2+\eta) = 2 \beta$ and $\hat{\eta} = 2\hat{\beta}
+ \hat{\nu} (d-2+\eta)$. From the standard scaling relations, the first of these
again recovers (\ref{Fi}).  (See also Ref.~\cite{Ab67Su68}.) 
From the scaling relations for logarithmic correction \cite{I} the second yields
$
 \hat{\eta} = d \hat{q} + \hat{\gamma} - \hat{\nu} (2 - \eta)
$.
When $\hat{q}$ vanishes, this is identical to (\ref{red}).
Indeed, this is the case in  the $d=2$, four-state Potts model \cite{I}
as well as 
in the $d=2$ pure and random Ising models \cite{Aade96,MaKu99,I}.
However, since ${\hat{q}} \ne 0$ at the upper critical dimension of 
the $O(N)$ $\phi^4$ theories and their long-range counterparts,
the percolation problem,   spin glasses and the Yang-Lee problem   \cite{Br82,I,Ru98},
this detailed matching of ${\cal{G}}_\infty(x,t)$ with $m_\infty^2(t)$  is invalid 
in these cases.
Instead, (\ref{red}) holds in each case.

In conclusion, then, the scaling theory presented in Ref.~\cite{I}
has been extended to deal with the specific heat when its leading exponent $\alpha$ vanishes
and the Fisher zeros impact onto the real axis at an angle other than $\pi/4$.
In such cases, (\ref{NewK}) is replaced by (\ref{NewK2}).
Also, the general theory has been extended to deal with the correlation function 
and the new relation (\ref{red}) has been checked against the literature and predictions made.
Together with Ref.~\cite{I}, the new formulae (\ref{NewK2}) and (\ref{red}) offer a set of scaling relations
analogous to the standard ones and appropriate to 
logarithmic corrections.

Besides these general results, progress specific to the random Ising models in two dimensions has been made.
Through (\ref{NewK2}) and (\ref{red}), the hitherto numerically most elusive and controversial quantity $\hat{\alpha}$
has been directly related to $\eta$, $\hat{\eta}$, $\hat{\gamma}$ and $\hat{q}$, all of which are clearly established.
Moreover, our theory automatically generates the famous double logarithm in the specific heat in these instances.

This work was partially supported by the EU RTN-Network ``ENRAGE'': {\em Random Geometry
and Random Matrices: From Quantum Gravity to Econophysics\/} under Grant~No.~MRTN-CT-2004-005616.

\vspace{-4mm}




\vspace*{-0.30cm}
\begin{thebibliography}{25}
\expandafter\ifx\csname natexlab\endcsname\relax\def\natexlab#1{#1}\fi
\expandafter\ifx\csname bibnamefont\endcsname\relax
  \def\bibnamefont#1{#1}\fi
\expandafter\ifx\csname bibfnamefont\endcsname\relax
  \def\bibfnamefont#1{#1}\fi
\expandafter\ifx\csname citenamefont\endcsname\relax
  \def\citenamefont#1{#1}\fi
\expandafter\ifx\csname url\endcsname\relax
  \def\url#1{\texttt{#1}}\fi
\expandafter\ifx\csname urlprefix\endcsname\relax\def\urlprefix{URL }\fi
\providecommand{\bibinfo}[2]{#2}
\providecommand{\eprint}[2][]{\url{#2}}

\vspace*{-0.30cm}

\bibitem{I} 
R.~Kenna, D.A.~Johnston, and W.~Janke, Phys. Rev. Lett. {\bf 96}, 115701 (2006).

\bibitem{history}
M.E.~Fisher, Rev. Mod. Phys. {\bf{70}}, 653 (1998).

\bibitem{ItDr89}
C.~Itzykson and J.M.~Drouffe, 
{\emph{Statistical Field Theory\/}}
(Cambridge University Press, 1989).

\bibitem{Wi65}
B.~Widom, J. Chem. Phys. {\bf{43}}, 3892 (1965); 
                                           {\bf{43}}, 3898 (1965).

\bibitem{Gr67}
R.B.~Griffiths, Phys. Rev. {\bf{158}}, 176 (1967).

\bibitem{Fi65} 
M.E.~Fisher, in {\emph{Lecture in Theoretical Physics VIIC\/}},
edited by W.E.~Brittin (University of Colorado Press, Boulder, 1965), p.~1.


\bibitem{FiSR64}
M.E.~Fisher, J.~Math.~Phys. {\bf{5}}, 944 (1964).

\bibitem{approp}
K.G.~Wilson, Phys. Rev. B {\bf{4}}, 3174 (1971);
                                           {\emph{ibid.\/}} 3184;
F.J.~Wegner,          {\emph{ibid.\/}}  4529 (1972);
in {\emph{Phase Transitions and Critical Phenomena\/}}, 
VI, ed. by C.~Domb and M.S.~Green (Academic Press, London, 1976), p.~8;
D.A.~Huse and M.E.~Fisher, J. Phys. C {\bf{15}}, L585 (1982);
A.~Aharony and M.E.~Fisher, Phys. Rev. B {\bf{27}}, 4394 (1983).

\bibitem{Aade96}
F.D.A.~Aar\~{a}o Reis, S.L.A.~de~Queiroz, and R.R.~dos~Santos, Phys. Rev. B {\bf{54}}, R9616 (1996); 
                                                         {\bf{56}},  6013 (1997). 

\bibitem{LaIg00}
P.~Lajk{\'{o}} and F.~Igl{\'{o}}i, Phys. Rev. E {\bf{61}}, 147 (2000).

\bibitem{SSLJ}
B.N.~Shalaev,  Sov. Phys. Solid State {\bf{26}},  1811 (1984);
               Phys. Rep.             {\bf{237}},  129 (1994);
R.~Shankar,    Phys. Rev. Lett.       {\bf{58}},  2466 (1987); 
                                      {\bf{61}},  2390 (1988);
A.W.W.~Ludwig, {\em ibid.\/}      {\bf{61}},  2388 (1988); 
               Nucl. Phys. B          {\bf{330}},  639 (1990);
G.~Jug and B.N.~Shalaev,    Phys. Rev. B {\bf{54}},  3442 (1996).




\bibitem{DD}
Vik.~S.~Dotsenko and Vl.~S.~Dotsenko, JETP Lett. {\bf{33}}, 37 (1981);
                                     Adv. Phys. {\bf{32}}, 129 (1983).


\bibitem{AnDo90WaSe90}
V.B.~Andreichenko, Vl.~S. Dotsenko, W.~Selke, and J.-S.~Wang, Nucl. Phys. B    {\bf{344}}, 531 (1990);
J.-S.~Wang, W.~Selke,  Vl.~S. Dotsenko, and  V.B.~Andreichenko, Europhys. Lett. {\bf{11}}, 301 (1990);
                                                                     Physica A {\bf{164}}, 221 (1990).

\bibitem{TaSh94WiDo95StAa97RoAd98}
A.L.~Talapov and L.N.~Shchur, J. Phys.: Condens. Matter {\bf{6}},   8295 (1994);  
S.~Wiseman and E.~Domany,                  Phys. Rev. E {\bf{51}},  3074 (1995);  
                                                        {\bf{52}},  3469 (1995); 
D.~Stauffer, F.D.A.~Aar{\~{a}}o~Reis, S.L.A.~de~Queiroz, and R.R.~dos~Santos,
                                   Int. J.  Mod. Phys. C {\bf{8}}, 1209 (1997); 
A.~Roder, J.~Adler, and W.~Janke,        Phys. Rev. Lett. {\bf{80}}, 4697 (1998);
                                       Physica A        {\bf{265}},  28 (1999).

\bibitem{BeSh04BeCh04}
B.~Berche and L.N.~Shchur, JETP Letters {\bf{79}}, 213 (2004);  
B.~Berche and C.~Chatelain,
in {\emph{Order, Disorder and Criticality\/}}, 
edited by Yu~Holovatch (World Scientific, Singapore, 2004), p.~146.

\bibitem{BaFe97SeSh98Pl98}
H.G.~Ballesteros, L.A.~Fern\'andez, V.~Mart\'{\i}n-Mayor, A.~Mu\~{n}oz~Sudupe,
G.~Parisi, and J.J.~Ruiz-Lorenzo,             J. Phys.~A {\bf{30}}, 8379 (1997);
W.~Selke, L.N.~Shchur, and O.A.~Vasilyev,      Physica~A {\bf{259}}, 388 (1998);
V.N.~Plechko,                              Phys. Lett.~A {\bf{239}}, 289 (1998).

\bibitem{Ki00}
J.-K~Kim, Phys. Rev. B {\bf{61}},  1246 (2000). 

\bibitem{finiteCinBond}
K.~Sawada and T.~Osawa,       Prog. Theor. Phys. {\bf{50}}, 1232 (1973);
T.~Tamaribuchi and F.~Takano,     {\emph{ibid.\/}} {\bf{64}}, 1212 (1980);
T.~Tamaribuchi,                             {\emph{ibid.\/}} {\bf{66}}, 1574 (1981).

\bibitem{ZieglerHe92}
K.~Ziegler, J.~Phys.~A         {\bf{18}}, L801  (1985); 
                               {\bf{21}}, L661 (1988); 
H.-O.~Heuer,  Phys. Rev. B {\bf{45}}, (1992) 5691. 

\bibitem{KiPa94Ku94KuMa00}
J.-K.~Kim and A.~Patrascioiu,  Phys. Rev. Lett. {\bf{72}}, 2785 (1994); 
                                                {\bf{73}}, 3489 (1994);
                               Phys. Rev. B     {\bf{49}}, 15764 (1994);
R.~K{\"{u}}hn,                Phys. Rev. Lett. {\bf{73}}, 2268 (1994); 
R.~K{\"{u}}hn and G.~Mazzeo, {\emph{ibid.\/}}  {\bf{84}}, 6135 (2000). 
 
\bibitem{Se94Zi94vaKu00}
W.~Selke, Phys. Rev. Lett. {\bf{73}}, 3487 (1994);
K.~Ziegler,                  \emph{ibid.\/} {\bf{73}}, 3488 (1994);
A.C.D.~van~Enter, C.~K{\"{u}}lske, and C.~Maes, {\emph{ibid.\/}} {\bf{84}}, 6134 (2000).  

\bibitem{KeLa91}
R.~Kenna and C.B.~Lang, Phys. Lett. B {\bf{264}}, 396 (1991);
                        Phys. Rev.~E {\bf{49}}, (1994) 5012;
                        Nucl.~Phys.~B {\bf{393}}, 461 (1993); 
                                      {\bf{411}}, 340 (1994);
R.~Kenna,            {\emph{ibid.\/}} {\bf{691}}, 292 (2004).

\bibitem{IPZ}
C.~Itzykson, R.B.~Pearson, and J.B.~Zuber,
Nucl. Phys. B {\bf{220}}, 415 (1983).


\bibitem{JaKe01}
W.~Janke and R.~Kenna, J. Stat. Phys. {\bf{102}}, 1211 (2001).

\bibitem{GPS}
A.~Caliri and D.C.~Mattis,                   Phys. Lett.~A {\bf{106}},  74, (1984);
M.L.~Glasser, V.~Privman, and L.S.~Schulman, J. Stat. Phys. {\bf{45}},  451, (1986);
                                              Phys. Rev.~B {\bf{35}}, 1841, (1987).

\bibitem{MaSh95}
V.~Matveev and R.~Shrock,        J. Phys.~A {\bf{28}}, 5235 (1995).

\bibitem{symm}
H.J.~Giacomini,                       Phys. Lett.~A  {\bf{115}}, 13  (1986);
R.~Kenna,                                     J. Phys.~A {\bf{31}}, 9419 (1998).

\bibitem{Fi78}
R.~Fisch, J. Stat. Phys. {\bf{18}}, 111 (1978).

\bibitem{GrWi81}
G.S.~Grest and M.~Widom, Phys. Rev. B {\bf{24}}, 6508 (1981).

\bibitem{SaSo97}
J. Salas and A.D. Sokal, J. Stat. Phys. {\bf{88}}, 567 (1997).

\bibitem{leadPNaSc80CaNa80}
B.~Nienhuis, E.K.~Riedel, and M.~Schick,          J. Phys. A  {\bf{13}},  L189 (1980);
J.L.~Cardy, M.~Nauenberg, and D.J.~Scalapino,     Phys. Rev. B {\bf{22}},  2560 (1980).

\bibitem{MaKu99} 
G.~Mazzeo and R. K{\"{u}}hn, Phys. Rev. E {\bf{60}}, 3823 (1999).  



\bibitem{moreetahat}
A.L.~Talapov and L.N.~Shchur, Europhys. Lett.      {\bf{27}},  193 (1994);  
S.L.A.~de~Queiroz and R.B.~Stinchcombe, Phys. Rev.~B {\bf{46}}, 6635 (1992);  
                                                                                          {\bf{50}}, 9976 (1994); 
                                        Phys. Rev.~E {\bf{54}}, 190  (1996); 
S.L.A.~de~Queiroz,                      {\emph{ibid.\/}} {\bf{51}}, 1030 (1995); 
                                          J. Phys.~A {\bf{30}}, L443 (1997); 
J.C.~Lessa and S.L.A.~de~Queiroz,                           cond-mat/0512407.  


\bibitem{BLZ}
E.~Br\'ezin, J.C.~Le~Guillou, and J.~Zinn-Justin, in {\emph{Phase Transitions and Critical Phenomena\/}},
VI, edited by D.~Domb and M.S.~Green (Academic Press, New York, 1976), p.~127.

\bibitem{FiMa72} 
M.E.~Fisher, S.-K.~Ma, and B.G.~Nickel, Phys. Rev. Lett. {\bf{29}}, 917 (1972).

\bibitem{LuBl97}
E.~Luijten and H.W.J.~Bl{\"{o}}te, Phys. Rev. B {\bf{56}}, 8945 (1997).

\bibitem{Pottslogs}
A.B.~Harris, J.C.~Lubensky, W.K.~Holcomb, and C.~Dasgupta, Phys. Rev. Lett. {\bf{35}}, 327 (1975);
I.W.~Essam,  D.S.~Gaunt, and A.J.~Guttmann,                J. Phys. A {\bf{11}}, 1983 (1978).

\bibitem{StJa03}
O. Stenull and H.K. Janssen, Phys. Rev. E {\bf{68}}, 036129  (2003).

\bibitem{spinglass} 
A.B.~Harris, T.C.~Lubensky, and J.-H.~Chen, Phys. Rev. Lett. {\bf{36}}, 415 (1976).

\bibitem{Ru98}
J.J.~Ruiz-Lorenzo, J.~Phys.~A {\bf{31}}, 8773 (1998).

\bibitem{Ab67Su68}
R.~Abe, Prog. Theor. Phys. {\bf 38}, 568 (1967);
M.~Suzuki, {\emph{ibid.\/}} {\bf 39}, 349 (1968).


\bibitem{Br82}
E.~Br\'ezin, J. Physique {\bf{43}}, 15 (1982).




\end{thebibliography}
\end{document}